\documentclass[12pt]{article}
\usepackage{fancyhdr}

\usepackage{mathrsfs}
\usepackage[T1]{fontenc}
\usepackage{mathpazo}
\usepackage{setspace}
\usepackage{amsfonts}
\usepackage{amssymb}
\usepackage{amsmath}
\usepackage{epsfig}
\usepackage{latexsym}
\usepackage{color}
\usepackage{graphicx}
\usepackage{nicefrac}
\usepackage[latin1]{inputenc}
\usepackage{slashed}
\usepackage{multirow}
\usepackage{fancybox}
\usepackage{cite}
\usepackage{comment}
\usepackage{soul}
\usepackage{color}

\usepackage[all,cmtip]{xy}
\usepackage{hyperref}

\def\hybrid{\topmargin -20pt    \oddsidemargin 0pt
        \headheight 0pt \headsep 0pt
        \textwidth 6.25in       
        \textheight 9.25in       
        \marginparwidth .875in
        \parskip 5pt plus 1pt   \jot = 1.5ex}

\hybrid

\def\baselinestretch{1.2}

\catcode`\@=11

\def\marginnote#1{}
\newcount\hour
\newcount\minute
\newtoks\amorpm
\hour=\time\divide\hour by60
\minute=\time{\multiply\hour by60 \global\advance\minute by-\hour}
\edef\standardtime{{\ifnum\hour<12 \global\amorpm={am}%
        \else\global\amorpm={pm}\advance\hour by-12 \fi
        \ifnum\hour=0 \hour=12 \fi
        \number\hour:\ifnum\minute<10 0\fi\number\minute\the\amorpm}}
\edef\militarytime{\number\hour:\ifnum\minute<10 0\fi\number\minute}

\def\draftlabel#1{{\@bsphack\if@filesw {\let\thepage\relax
   \xdef\@gtempa{\write\@auxout{\string
      \newlabel{#1}{{\@currentlabel}{\thepage}}}}}\@gtempa
   \if@nobreak \ifvmode\nobreak\fi\fi\fi\@esphack}
        \gdef\@eqnlabel{#1}}
\def\@eqnlabel{}
\def\@vacuum{}
\def\draftmarginnote#1{\marginpar{\raggedright\scriptsize\tt#1}}

\def\draft{\oddsidemargin -.5truein
        \def\@oddfoot{\sl preliminary draft \hfil
        \rm\thepage\hfil\sl\today\quad\militarytime}
        \let\@evenfoot\@oddfoot \overfullrule 3pt
        \let\label=\draftlabel
        \let\marginnote=\draftmarginnote
   \def\@eqnnum{(\theequation)\rlap{\kern\marginparsep\tt\@eqnlabel}%
\global\let\@eqnlabel\@vacuum}  }

\def\preprint{\twocolumn\sloppy\flushbottom\parindent 2em
        \leftmargini 2em\leftmarginv .5em\leftmarginvi .5em
        \oddsidemargin -.5in    \evensidemargin -.5in
        \columnsep .4in \footheight 0pt
        \textwidth 10.in        \topmargin  -.4in
        \headheight 12pt \topskip .4in
        \textheight 6.9in \footskip 0pt
        \def\@oddhead{\thepage\hfil\addtocounter{page}{1}\thepage}
        \let\@evenhead\@oddhead \def\@oddfoot{} \def\@evenfoot{} }

\def\numberbysection{\@addtoreset{equation}{section}
        \def\theequation{\thesection.\arabic{equation}}}

\def\underline#1{\relax\ifmmode\@@underline#1\else
        $\@@underline{\hbox{#1}}$\relax\fi}

\def\titlepage{\@restonecolfalse\if@twocolumn\@restonecoltrue\onecolumn
     \else \newpage \fi \thispagestyle{empty}\c@page\z@
        \def\thefootnote{\fnsymbol{footnote}} }

\def\endtitlepage{\if@restonecol\twocolumn \else \newpage \fi
        \def\thefootnote{\arabic{footnote}}
        \setcounter{footnote}{0}}  

\catcode`@=12
\relax

\def\figcap{\section*{Figure Captions\markboth
        {FIGURECAPTIONS}{FIGURECAPTIONS}}\list
        {Figure \arabic{enumi}:\hfill}{\settowidth\labelwidth{Figure
999:}
        \leftmargin\labelwidth
        \advance\leftmargin\labelsep\usecounter{enumi}}}
 \relax
\def\tablecap{\section*{Table Captions\markboth
        {TABLECAPTIONS}{TABLECAPTIONS}}\list
        {Table \arabic{enumi}:\hfill}{\settowidth\labelwidth{Table
999:}
        \leftmargin\labelwidth
        \advance\leftmargin\labelsep\usecounter{enumi}}}
 \relax
\def\reflist{\section*{References\markboth
        {REFLIST}{REFLIST}}\list
        {[\arabic{enumi}]\hfill}{\settowidth\labelwidth{[999]}
        \leftmargin\labelwidth
        \advance\leftmargin\labelsep\usecounter{enumi}}}
 \relax
\makeatletter
\newcounter{pubctr}
\def\publist{\@ifnextchar[{\@publist}{\@@publist}}
\def\@publist[#1]{\list
        {[\arabic{pubctr}]\hfill}{\settowidth\labelwidth{[999]}
        \leftmargin\labelwidth
        \advance\leftmargin\labelsep
        \@nmbrlisttrue\def\@listctr{pubctr}
        \setcounter{pubctr}{#1}\addtocounter{pubctr}{-1}}}
\def\@@publist{\list
        {[\arabic{pubctr}]\hfill}{\settowidth\labelwidth{[999]}
        \leftmargin\labelwidth
        \advance\leftmargin\labelsep
        \@nmbrlisttrue\def\@listctr{pubctr}}}
 \relax
\makeatother
\newskip\humongous \humongous=0pt plus 1000pt minus 1000pt

\newif\ifdtup

\relax

\def\be{\begin{equation}}
\def\ee{\end{equation}}
\def\ba{\begin{eqnarray}}
\def\ea{\end{eqnarray}}

\def\del{\partial}

\def\k{\kappa}
\def\r{\rho}
\def\a{\alpha}

\def\b{\beta}

\def\g{\gamma}
\def\G{\Gamma}
\def\d{\delta}
\def\D{\Delta}
\def\e{\epsilon}

\def\C{\Chi}

\def\m{\mu}
\def\n{\nu}
\def\om{\omega}
\def\Om{\Omega}
\def\l{\lambda}
\def\L{\Lambda}
\def\s{\sigma}
\def\S{\Sigma}

\def\no{\noindent}

\def\qq{\qquad}

\def\IR{\relax{\rm I\kern-.18em R}}

\def \ha {{1\over 2}}

\def \ov {\over}

\def\IR{\relax{\rm I\kern-.18em R}}
\def\IL{\relax{\rm I\kern-.18em L}}

\def\inv{^{\raise.15ex\hbox{${\scriptscriptstyle -}$}\kern-.05em 1}}

\begin{document}

\renewcommand{\theequation}{\thesection.\arabic{equation}}
\csname @addtoreset\endcsname{equation}{section}

\newcommand{\beq}{\begin{equation}}
\newcommand{\eeq}[1]{\label{#1}\end{equation}}
\newcommand{\ber}{\begin{eqnarray}}
\newcommand{\eer}[1]{\label{#1}\end{eqnarray}}
\newcommand{\eqn}[1]{(\ref{#1})}
\begin{titlepage}
\begin{center}


~

\vskip 1 cm

{\large \bf  Spacetimes for  $\lambda$-deformations  }

\vskip 0.4in

{\bf Konstantinos Sfetsos}$^1$\phantom{x} and\phantom{x} {\bf Daniel C. Thompson}$^2$\vskip 0.1in
{\em
${}^1$Department of Nuclear and Particle Physics\\
Faculty of Physics, University of Athens\\
Athens 15784, Greece\\
 {\tt \footnotesize ksfetsos@phys.uoa.gr}}
\vskip
0.1in
{\it ${}^2$ Theoretische Natuurkunde, Vrije Universiteit Brussel,
and The International Solvay Institutes,
Pleinlaan 2, B-1050 Brussels, Belgium.\\
{\tt \footnotesize daniel.thompson@vub.ac.be}}
\vskip .5in
\end{center}

\centerline{\bf Abstract}
\no
We examine a recently proposed class of integrable deformations to
two-dimensional conformal field theories. These $\lambda$-deformations interpolate
between a WZW model and the non-Abelian T-dual of a Principal Chiral Model on a group $G$ or,
between a $G/H$ gauged WZW model and the non-Abelian T-dual of the geometric coset $G/H$.
$\l$-deformations have been conjectured to represent quantum group $q$-deformations for the
case where the deformation parameter is a root of unity.
In this work we show how such deformations can be given an embedding as full string backgrounds
whose target spaces satisfy the equations of type-II supergravity.
One illustrative example is a deformation of the $Sl(2,\IR)/U(1)$ black-hole CFT.
A further example interpolates between the $\frac{SU(2)\times SU(2)}{SU(2)} \times   \frac{SL(2,\IR)\times SL(2,\IR)}{SL(2,\IR)}
\times U(1)^4$ gauged WZW model
and the non-Abelian T-dual of $AdS_3\times S^3 \times T^4$ supported with Ramond flux.

\no

\newpage

\tableofcontents

\noindent

\vskip .4in

\end{titlepage}
\vfill
\eject

\def\baselinestretch{1.2}
\baselineskip 20 pt
\noindent


\section{Introduction}

The Wess-Zumino-Witten (WZW) model \cite{Witten:1983ar}
 and the Principal Chiral Model (PCM) \cite{Polyakov:1975rr} for a group manifold
$G$ provide two of the most well studied examples of two-dimensional integrable systems and are of immense importance in
many areas of theoretical and mathematical physics.
Key to their simplicity is the underlying group structure;  the WZW model is a current algebra Conformal Field Theory (CFT), whereas
 the equations  of motion and Bianchi identities for
the currents of the PCM can be combined into a Lax equation for a Lax connection from which an infinite number of conserved quantities can be deduced
\cite{Luscher:1977rq}.
It is natural to ask whether it is possible to deform such theories whilst preserving integrability.
It was observed in \cite{Rajeev:1988hq}
that there exists a one-parameter deformation of the  canonical Poisson structure of the PCM which defines two commuting Kac-Moody alegbras
and preserves integrablity.  Some years later, a first step to finding a Lagrangian description of these deformed theories was made in \cite{Balog:1993es}
for the case where the underlying group $G=SU(2)$.
Due to technical complexity involved, extending the direct approach of  \cite{Balog:1993es}  to arbitrary groups seemed rather intractable.

However, very recently  \cite{Sfetsos:2013wia} the Lagrangian description of the deformed theories for any group $G$ was provided. The approach of \cite{Sfetsos:2013wia} was to consider a total action comprised of
the sum of PCM parametrized by group element  $\tilde{g}\in G$ together with a WZW  parametrized by a $g \in G$.
The combined action enjoys a $G_L \times G_R$ global symmetry of the PCM and a $G_{L,{\rm cur}}\times G_{R,{\rm cur}}$ current symmetry of the WZW.
The critical step is to then gauge  a subgroup of the global symmetry   that acts as $G_L$ on $\tilde g$ and $G_{\rm diag}$ on $g$.
The gauge symmetry can be fixed, for instance by setting $\tilde{g}=\mathbb{1}$, and the non-propagating gauge fields may be integrated out.
The result is a $\sigma$-model that depends on the level $k$ of the WZW  and the `radius' $\k^2$  of the PCM in the combination
\be
\label{eq:lambda}
\lambda  = \frac{k}{k + \kappa^2} \ ,
\ee
which can be related to  the deformation parameter of \cite{Rajeev:1988hq,Balog:1993es}.

When the level of the WZW is much smaller than the radius of the PCM, the later is effectively frozen out of the dynamics.
Indeed, for small $\lambda$ the result is to deform the WZW CFT by a current-current bilinear.

The opposite limit  $\lambda \rightarrow 1$ requires more care;  it was shown in  \cite{Sfetsos:2013wia} that if the
group element $g$ is appropriately expanded near the identity the result is to produce a $\sigma$-model whose spacetime is the non-Abelian T-dual
of the PCM on the group space $G$ with respect to the $G_L$ action.  So for $\lambda$ near unity, one can also view this as a regulated version
of non-Abelian T-dual resolving global ambiguities.\footnote{This idea originated in \cite{Sfetsos:1994vz}. It was more recently put forward and further tested in \cite{Polychronakos:2010hd}.}

By either a direct calculation of the algebra of non-local charges or via an expansion of the Maillet-type
Poisson brackets for the monodromy matrix, these theories can be seen to exhibit the whole Yangian symmetry for all
values of the deformation $0\leqslant \l \leqslant 1$ \cite{Itsios:2014vfa}.

With some modification, this general construction of integrable deformations can be applied
to strings on cosets or symmetric spaces giving rise to integrable deformations of coset CFTs \cite{Sfetsos:2013wia,Hollowood:2014rla}.
Furthermore it has been extended, with obvious applications to AdS/CFT, to Green-Schwarz superstrings  on super-cosets \cite{Hollowood:2014qma}.
In this later context one may make connections with other known deformations of superstrings in $AdS_{5}\times S^{5}$.
A sequence of works \cite{Hoare:2011nd,Hoare:2011wr,Hoare:2012fc,Hoare:2013ysa} have studied how the symmetries of the world sheet S-matrix  may be deformed to a quantum group
whilst still satisfying S-matrix axioms.  The deformation is labeled by a parameter $q$ and there are two cases to consider.
First is $q = e^{\eta} \in \mathbb{R}$  which corresponds to the ``$\eta$-deformation'' introduced from the string world sheet
perspective in \cite{Delduc:2013fga,Delduc:2013qra} building on earlier work in \cite{Klimcik:2002zj}.  The $\eta$-deformation   has been further developed in \cite{Arutyunov:2013ega,Hoare:2014pna,Arutynov:2014ota}.  The second case is when $q$ is a root of unity and it was conjectured in  \cite{Hollowood:2014qma} that the $\lambda$-deformed
theories described above give a world sheet realisation for this scenario.

A crucial question is then  whether these integrable deformations
are marginal and thus give rise to a target space that is a consistent string theory background.
Working in a $\kappa$-fixed Green-Schwarz style action makes it technical to ascertain the full geometry of such a deformation.
Within the context of  just bosonic string theory the   deformation of the WZW CFT fixed point
is certainly {\em not} marginal according to the results of \cite{Chaudhuri:1988qb}  (indeed the running of $\lambda$
was calculated in \cite{Itsios:2014lca,Sfetsos:2014jfa} and shown to agree with that of the non-Abelian bosonized Thirring model
computed in \cite{Gerganov:2000mt}).
Further fermionic field content, coming from the RR sector of the type-II superstring, is needed so that one-loop conformal invariance is
preserved for all values of $\lambda$. This then provides the motivation to the question we address here: can we embed the
target spaces corresponding to integrable $\lambda$-deformations as full solutions of type-II supergravity?

We will show, with a number of worked examples, that this is indeed possible.
It is, by no means, obvious that this will be the case; indeed a number of simplistic first attempts at this problem yielded no success.
Our results come from two observations.    First  applying the above deformation to a compact group will give a $\lambda$
dependent positive contribution to the one-loop dilaton beta-function.  To counter balance this it seems necessary to perform
a similar $\lambda$ deformation in a non-compact group.  Second, at the $\lambda = 1$ fixed point
which we recall is the $G_L$ non-Abelian T-dual of a PCM, we can embed the geometry into a solution of supergravity
by the inclusion of Ramond fields determined by group theoretic considerations in \cite{Sfetsos:2010uq}.
The close relation to non-Abelian T-duality suggests that the techniques of \cite{Sfetsos:2010uq} may be generalised
to find appropriate supporting Ramond fluxes for all values of $\lambda$.

In this paper we will explicitly consider examples of $\lambda$-deformations applied to :
\begin{enumerate}
\item   $AdS_3\times S^3$   using the $SU(2)\times SL(2,\IR)$ isometry of the group
\item   $AdS_2 \times S^2$ using the $SU(2)\times SL(2,\IR)$ isometry of the maximal coset
\item  $AdS_3\times S^3$   using the $SU(2)\times SU(2) \times SL(2,\IR)\times SL(2,\IR)$ isometry of the maximal coset
\end{enumerate}
The  first example is somewhat simpler since it uses groups rather than cosets and we
include it for didactic purposes. However, in this case one is forced to the conclusion that the RR fields
must be imaginary and thus constitute a solution of type-II$^\star$ rather than type-II theories \cite{Hull:1998vg}.
The reader who is dissatisfied with this state of affairs should quickly move to the second and third example which employ
the coset generalisation of \cite{Sfetsos:2013wia} -- though a little more involved, these are real backgrounds of type-II theories.

The rest of the paper is as follows:  We begin in section \ref{frame} by reviewing the general construction of  \cite{Sfetsos:2013wia} for group manifolds.
In section \ref{sec:examplegroup} we provide the first of the examples listed above including a high level of detail and methodology.
In section \ref{sec:cosets} we describe the generalisation to cosets and follow this with the remaining examples in section    \ref{sec:5}
and section \ref{sec:example3}.

\section{ $\lambda$-deformations for groups }
\label{frame}

In this section we present the background fields for the NS-sector of our models.
In order to set up our notation and make our paper self-contained we first briefly review the necessary results and conventions.

Consider a general compact group $G$ and a corresponding group element
$g$ parametrized by $X^\m$, $\m=1,2,\dots , \dim(G)$. The right and
left invariant Maurer--Cartan forms, as well as the orthogonal
matrix (or adjoint action) relating them, are defined as
\be
\begin{aligned}
 & J^A_+ = -i\, {\rm Tr}(T^A \del_+ g g^{-1}) = R^A_\m \del_+ X^\m \ ,  \qq J^A_- = -i\, {\rm Tr}(T^A g^{-1} \del_- g )= L^A_\m \del_- X^\m\ ,
\\
 & R^A_\m = D_{AB}L^B_\m\ ,  \qq D_{AB}={\rm Tr}(T_A g T_B g^{-1})\ .
 \label{jjd}
 \end{aligned}
\ee
The matrices $T^A$ obey $[T_A,T_B]=i f_{ABC} T_C$ and are normalized as ${\rm Tr}(T_A T_B)=\d_{AB}$.

\no
The PCM on the group manifold for an element $\tilde{g}\in G$ is
\be
\label{eq:PCM}
S_{{\rm PCM}}(\tilde{g}) = \frac{\k^{2}}{\pi} \int_{\Sigma} \d^{AB} L_{+}^{A}(\tilde{g})  L_{-}^{B}(\tilde{g})
\ee
and enjoys a $G_{L}\times G_{R}$ global symmetry.  The WZW action for a group element $g  \in G$  is defined by

\be
\label{eq:WZW}
S_{{\rm WZW},k}(g) =
\frac{k}{2\pi} \int_{\Sigma} \delta_{AB}L^A_+(g) L^B_-(g) \,
  +\frac{k}{12\pi}\int_{{\cal B}} f_{ABC} L^A\wedge
L^B\wedge L^C\ ,
\ee
where ${\cal B}$ is an extension such that $\partial {\cal B} = \Sigma$.
The approach of \cite{Sfetsos:2013wia} was to consider the sum of the actions in \eqref{eq:PCM} and \eqref{eq:WZW}
and to gauge a subgroup of the global symmetries that acts as
\be
\tilde{g} \rightarrow h^{{-1}}\tilde{g} \ , \qq g \rightarrow h^{-1} g h \ , \qq h \in G \ .
\ee
This is achieved by introducing a connection $A= A^{A}T^{A}$ valued in the alebgra of $G$ that transforms as
\be
A \rightarrow   h^{-1} A h - h^{-1} dh\ , \qq h \in G \ .
\ee
We  replace derivatives in the PCM with covariant derivatives defined as
\be
D \tilde{g}  =  d \tilde{g} -A\tilde{g}
\ee
and replace the $WZW$ with the $G/G$ gauged $WZW$ given by
\be
S_{{\rm gWZW},k}(g, A)= S_{{\rm WZW},k}(g) +   {k\ov \pi} \int {\rm Tr}(A_-\del_+ gg^{-1} - A_+ g^{-1} \del_-  g + A_- g A_+ g^{-1}- A_-A_+) \ .
\ee
The gauge symmetry can now be gauged fixed by setting $\tilde{g}= \mathbb{1}$
such that all that remains of the gauged PCM is a quadratic term in the gauge fields.
The gauge fields, which are non-propagating are integrated out to result in the $\s$-model action
\cite{Sfetsos:2013wia}
\be
S_{k,\l}(g) =
S_{{\rm WZW},k}(g) + {{k}\ov \pi}  \int J_+^A
(\l^{-1}-D^T)^{-1}_{AB}J_-^B  \ ,
\label{tdulalmorev2}
\ee
where
\be
\l _{AB}= \lambda \d_{AB}\ ,\qq \l = {k\ov k+\k^2}\ .
\label{lee}
\ee
The $\s$-model of \eqn{tdulalmorev2} is integrable as was  proven in \cite{Sfetsos:2013wia} by showing that the
corresponding metric and antisymmetric tensor fields satisfy the
algebraic constraints for integrability of \cite{Balog:1993es} and
\cite{Evans:1994hi}.  A form of the action similar to \eqn{tdulalmorev2} appeared before in
\cite{Tseytlin:1993hm}.

We note that a more general class of actions can be obtained by retaining $\l_{AB}$
as a general constant matrix (or one that depends only on spectator fields and not the $X^{\mu}$).
Such models are obtained by repeating the same procedure but replacing the inner product $\k^2 \d_{AB}$ occurring in
the PCM of \eqref{eq:PCM} with a general constant coupling matrix $E_{AB}$ to which $\l_{AB}$ is
related with a straightforward extension of \eqn{lee} \cite{Sfetsos:2013wia,Sfetsos:2014jfa}.
It remains an open question as to which choices $E_{AB}$ can be made whilst retaining integrability.
With future possibilities in mind many of our derivations are done keeping $\l_{AB}$ as a general matrix.

\subsection{Limit properties}

In the limit of small $\l_{AB}$ the action  \eqn{tdulalmorev2}
can be approximated by
\be
S_{k,\l}(g) = S_{{\rm WZW},k}(g) + {k\ov \pi} \int \l_{AB} J_+^AJ_-^B +{\cal O}(\l^2)\ ,
\label{thorio}
\ee
corresponding to the WZW theory perturbed by the current bilinear $J_+^AJ_-^B$ with arbitrary coupling matrix $\l_{AB}$.
The first two terms define the bosonized anisotropic non-Abelian Thirring model in analogy
with the non-Abelian Thirring model \cite{Dashen:1974gu,Karabali:1988sz}.
Hence, it is reasonable to expect that \eqn{tdulalmorev2} provides an effective all loop action for the
bosonized non-Abelian Thirring model. Based on studies of the RG flow and symmetry considerations
this has been shown for $\l_{AB}=\l \d_{AB}$ in \cite{Itsios:2014lca} and for general $\l_{AB}$ in \cite{Sfetsos:2014jfa}.
The fact that the model is driven away from the conformal point by the current bilinear in
\eqn{thorio} is tied to the non-Abelian nature of the group. Had the current bilinear been restricted to the Cartan torus, as
for instance in \cite{Hassan:1992gi},  the $\s$-model would have remained conformal \cite{Chaudhuri:1988qb}.   We remark that other interesting types of marginal deformations have included the so-called asymmetric deformations of the form $\int d^2z J  \bar{J}_G$ in which $\bar{J}_G$  corresponds to some other $U(1)$ outside the chiral ring of the WZW \cite{Kiritsis:1994ta,Kiritsis:1995iu}.  Heterotic embeddings of these asymmetric deformations were considered in \cite{Israel:2004vv}.

\no
What motivated the present paper is the behaviour of \eqn{tdulalmorev2} for $k\gg 1$ and $\l\to \mathbb{1}$ \cite{Sfetsos:2013wia}. Then
expanding the matrix and group elements near the identity we have that
\be
\l_{AB} =\d_{AB} - {1\ov k} E_{AB} + {\cal O}\left(1\ov k^2\right)\ , \quad g = \mathbb{1} + i { v_A T^A \ov k} + {\cal O}\left( 1\ov k^2 \right)  \ ,
\label{laborio}
\ee
 leading to
\be
J_\pm^A = {\del_\pm v^{A}\ov k} + {\cal O}\left( 1\ov k^2 \right)  \ ,\quad
D_{AB} = \d_{AB} + {f_{AB} \ov k} + {\cal O}\left( 1\ov k^2 \right) \ ,\quad f_{AB} = f_{ABC} v_C\ .
\ee
In this limit the action \eqn{tdulalmorev2} becomes
\be
S_{\rm non\!-\!Abel}(v) = {1\ov \pi} \int \del_+ v^A (E + f)^{-1}_{AB}\del_- v^B  \ ,
\label{nobag}
\ee
which is the the non-Abelian T-dual with respect to the $G_{L}$ action of the $\s$-model given by  the PCM action with general coupling matrix $E_{AB}$.

\subsection{Towards a supergravity embedding }

The purpose of this paper is to embed models for which the metric and antisymmetric tensor of the NS sector are provided by \eqn{tdulalmorev2}
to type--II supergravity. This will be done by supporting these fields with a dilaton as well as with appropriate RR fluxes.
The dilaton factor is obtained from integrating out the gauge fields in a path integral and is given by
\be
e^{-2\Phi} = e^{-2 \Phi_0} k^{\dim G} \det (\l^{-1} - D^T) \ ,
\ee
where $\Phi_0$ is the dilaton of the original theory in which the PCM is a part of.
 \no
Let us now restrict to the simplest cases when $\l_{AB} = \l \d_{AB}$.
 The target space metric can be read from the $\sigma$-model \eqn{tdulalmorev2}  and can be conveniently expressed using frame fields

two frames $e_+^A$ and $e_-^A$ given by \cite{Sfetsos:2013wia,Itsios:2014lca}
\begin{equation}
\label{eq:frames1}
e_+^A = -\sqrt{k(1-\l^2)}(D - \lambda \mathbb{1} )^{-1}_{AB} R^B \ , \qq
e_-^a = \sqrt{k(1-\l^2)} (D^T- \lambda  \mathbb{1})^{-1}_{AB} L^B\ .
\end{equation}
Both these frame field define the same geometry and are related according to a local frame rotation $e_- = \Lambda e_+$  given by
\begin{equation}
\Lambda  =  - ( \mathbb{1} - \lambda D)^{-1} (D-\lambda \mathbb{1}) \ .
\label{llore}
\end{equation}

 In the spirit of  \cite{Sfetsos:2010uq} we will need the orthogonal transformation in the spinor representation
$\Om$ found from
\be
\Om^{-1} \G^A \Om = \L^A{}_B \G^B\ ,
\label{dsjfh}
\ee
where $\G^A$ are the ten-dimensional $\G$-matrices.
An ansatz  for the RR fields, completely determined by the group theory,
is  given by allowing this $\Omega$ matrix to act by Clifford multiplication on the RR fields
of the original model in much the same as it does for both Abelian and non-Abelian T-duality.
We consider the RR sector in the democratic formalism that incorporates fluxes and their Hodge duals equally specified by polyforms
 \be\label{RRpoly}
{\rm IIB}:\   {\mathbb{F}}   =  \sum_{n=0}^4  F_{2n+1}\ ,
\qquad
{\rm IIA}:\   \widehat{\mathbb{F}}  =  \sum_{n=0}^5 F_{2n}\ ,
\ee
from which one obtains bi-spinors  $\slashed{\mathbb{F}}$  by contracting the constituent
p-forms with p-anti-symmetrised gamma matrices.  Then the ansatz we propose for the RR sector is
 \be
 \label{RRrule}
 e^{\Phi}   \slashed{\mathbb{F}}   =  \mu(\lambda) e^{\Phi_{0}} \slashed{\mathbb{F}}_{0} \cdot \Omega^{-1}  \ ,
 \ee
 where on the right hand side we have the bispinor $\slashed{\mathbb{F}}_{0}$ formed from the RR fields
supporting the PCM of \eqref{eq:PCM} when embedded into a supergravity background.
The RR fields across the deformation are obtained from the left hand side of this relation
upon replacing the anti-symmetrised gamma matrices with the wedge of the corresponding frame fields $e^{A}$ given in \eqref{eq:frames1}.
We include in this  ansatz a possible multiplicative $\lambda$ dependent constant coefficient $\m(\l)$ which is,
of course, related to the normalisation employed for the dilaton. This constant should necessarily vanish
in  the limit $\l\to 0$ since then the background, being a CFT, should consist of purely NS fields.
We believe this to be the only consistent ansatz for the RR fields which is compatible with the group theoretic
structure of the problem and  that agrees with the result in the non-Abelian T-dual limit.
A first principles derivation of the form of the RR fields should eventually be established  using e.g. a Green Schwarz or pure-spinor formalism.

\section{Integrable deformation based on $SU(2)$ and $SL(2,\IR)$}\label{sec:examplegroup}

With the aim of constructing an integrable deformation of $AdS_3\times S^3\times T^4$
we recap the application  of the above construction for the case of the $SU(2)$ group manifold
given in  \cite{Sfetsos:2013wia} for the NS sector.  We then provide the analytic continuation that
gives an analogous result for $SL(2,\IR)$ before combining these to give a full supergravity embedding.
To prevent feelings of resentment on behalf of the reader, let us state the outcome upfront: in this case
the result will be a background of Type-IIB$^{\star}$ supergravity with imaginary fluxes.
To obtain real backgrounds of type-II supergravity we will consider the generalisation of the these $\lambda$ deformations
to coset spaces in later sections.

\subsection{The $SU(2)$ integrable deformation}\label{sec:Su2def}

We parametrize an $SU(2)$ group element by
\be
g = e^{i \a \hat n_i \s_i}\ ,\qq {\bf \hat n} = (- \sin\b \sin\g, \sin\b \cos\g,  \cos\b)\ ,
\ee
where $\s_i$'s are the Pauli matrices such that
\be
 g=\left(
                             \begin{array}{cc}
                              a_0+ i a_3 &  a_2 + i a_1 \\
                                -a_2 + i a_1 & a_0 - ia_3 \\
                             \end{array}
                          \right) =
\left(
   \begin{array}{cc}
     \cos{\a} + i \sin\a \cos\b & \sin\a \sin\b \ e^{-i\g} \\
     -\sin\a \sin\b \ e^{i\g} & \cos{\a} - i \sin\a \cos\b \\
   \end{array}
 \right)\ .
\label{su2g2}
\ee
The corresponding $\s$-model has metric and NS two-form
\be
\begin{split}
\hskip -2 cm
SU(2): \qq
&ds^2 = k \left({1+\l\ov 1-\l} d\alpha^2 + {1-\l^2\ov \D(\a)} \sin^2\alpha \ ds^2(S^2) \right)\ ,
\\
& B = k\left( - \alpha + \frac{(1-\lambda)^2}{\D(\a)} \cos\a \sin \alpha \right)  {\rm{Vol}}(S^2)\ ,
\end{split}
\label{su2def}
\ee
where we have defined
\be
\D(\a) = (1-\l)^2 \cos^2\a + (1+\l)^2 \sin^2 \a
\ee
and $ds^2(S^2) = d\b^2 + \sin^2\!\b\ d\g^2$ and ${\rm{Vol}}(S^2) =\sin\b\ d\b \wedge d\g$.
Note that for $\a\to 0$ and for $\a\to \pi$ the geometry becomes $\mathbb{R}^3$. For $\a\to \pi/2$ it becomes
$S^1\times S^2$.

\no
The Lorentz rotation matrix define by \eqn{llore} is given by
\ba
&& \L_{AB} = {1\ov \D(\a)} \Big[((1+\l)^2\sin^2\a - (1-\l)^2 \cos^2\a)\d_{AB}
\nonumber\\
&& \phantom{xxxx} -2 (1-\l^2)\cos\a\ \e_{ABC} a_C - 2 (1+\l)^2 a_A a_B\Big]\ ,
\ea
and its spinor representation \eqn{dsjfh} is given by
\be
\Om = {1\ov \sqrt{\D(\a)}} \G_{11} \big[ (\l-1)\cos\a \G_{123} + (1+\l)\sin \a\ {\bf \hat n}\cdot {\bf \G}\big]\ .
\label{djkhll}
\ee

\no
In the limit $\l\to 0$ we obtain the metric and the antisymmetric tensor for
the $SU(2)$ WZW model with normalization such that $R_{ij} =2 g_{ij}$. In this case
the rotation matrix in the spinor representation \eqn{djkhll} becomes
\be\label{eq:OmegaSU2}
\Omega =\G_{11}\big( -\cos \a \G_{123} + \sin \a\ {\bf \hat n}\cdot {\bf \G}\big)\ .
\ee
The fact that it is not the identity is in agreement with the fact that the frames become in that limit
$e_+^A = - R^A$ and $e_-^A = L^A$. This implies that in this limit $\L_{AB}= -D_{AB}$, consistent with \eqn{llore}.

\no
The non-Abelian limit of the $SU(2)$ PCM  is obtained (setting $\k =1$) by letting
\be
\a = {r\ov 2k} \ ,\qq \l= {k\ov 1+k} \ ,\qq k\to \infty\ .
\label{alim}
\ee
This limiting procedure gives
\be
ds^2 =\ha \left(dr^2 + {r^2\ov r^2+1} ds^2(S^2)\right) \ ,\qq
B = -\ha {r^3\ov r^2+1} {\rm Vol}(S^2)\ .
\label{dsbfg}
\ee
In addition the Lorentz transformation in the spinor representation \eqn{djkhll} becomes
\be
\Om = {\G_{11}\ov \sqrt{1+r^2}} (-\G_{123} + {\bf v}\cdot  {\bf \G})\ ,
\ee
where ${\bf v}= r {\bf \hat n}$.
These expressions correspond the non-Abelian T-dual of the $SU(2)$ PCM which in fact has been embedded in supergravity.
It was shown in \cite{Sfetsos:2010uq} that when supported by appropriate flux fields it is a solution of massive IIA-supergravity
and that it represent the non-Abelian T-dual of the background corresponding to the near horizon of the D1-D5 brane system.

\subsection{The $SL(2,\IR)$ integrable deformation}

The case with $G=SL(2,\IR)$ can be obtained by an analytic continuation
$\b\to i \tilde \b$ and the simultaneous flip of sign of $k$. In addition we rename $\a\to \tilde \a$ and $\g\to \tilde \g$.
Then from \eqn{su2def} we obtain the background
\be
\begin{split}
\hskip -2 cm
SL(2,\IR): \qq &ds^2 = k \left(-{1+\l\ov 1-\l} d\tilde \alpha^2 + {1-\l^2\ov \D(\tilde \a)} \sin^2\tilde \alpha \ ds^2(H^2) \right)\ ,
\\
& B = k\left( - \tilde \alpha + \frac{(1-\lambda)^2}{\D(\tilde \a)} \cos\tilde \a \sin \tilde \alpha \right)  {\rm{Vol}}(H^2)\ ,
\end{split}
\label{sl2def}
\ee
where $ds^2(H^2) = d\tilde\b^2 + \sinh^2\!\tilde\b\ d\tilde\g^2$ and ${\rm{Vol}}(H^2) =\sinh \tilde\b\ d\tilde\b \wedge d\tilde\g$.
As expected the metric has signature $(-,+,+)$.

\subsection{Embedding to supergravity}

We consider performing the above procedure on $SL(2,\IR)\times SU(2)\times T^4$.
i.e. we look for an integrable deformation to the model whose target
space is $AdS_3 \times S^3\times T^4$ supported in  type-IIB supergravity by   an RR three-form flux field
\be
F_3 = \sqrt{2} ( e^0\wedge e^1\wedge e^2 + e^3\wedge e^4\wedge e^5)\ ,
\label{frr3}
\ee
where the indices $0,1,2$ and $3,4,5$ run along the $AdS_3$ and $S^3$
directions.\footnote{We normalize the metric so that $R_{\m\n}=\mp g_{\m\n}$ for
 the $AdS_3$ (upper sign) and the $S^3$ (lower) and set the dilaton to be zero.
We are not considering at this stage the $(p,q)$ string case where the geometry is supported by both NS and RR flux in combination.}

Turning to the deformation we note that this will be non-trivial for the $SU(2)$ and $SL(2,\IR)$ factors as presented in the previous section.
There will be no deformation for $T^4$. The metric and the NS antisymmetric tensor are given by
\be
ds^2 =  ds^2_{SL(2,\IR),\l} + ds^2_{SU(2),\l}  + \sum_{i=1}^4 dx_i^2  \
\ee
and
\be
B =  B_{SL(2,\IR),\l} + B_{SU(2),\l}  \ ,
\ee
with the obvious notation for the various terms corresponding to \eqn{su2def} and \eqn{sl2def}.
The geometry will supported by a dilaton field given by
\be
e^{-2\Phi} = \D(\a)\D(\tilde \a)\ .
\ee
 The frame we will use are given by
\ba
\begin{split}
& \frak{e}^{0} = \sqrt{k\frac{1+\lambda }{1-\lambda} }
d\tilde \alpha  \ , \quad \frak{e}^{1} =
\sqrt{k\frac{1-\lambda^2}{  \D(\tilde \alpha) } }\sin\tilde\alpha
d\tilde\beta  \ ,  \quad
\frak{e}^{2} = \sqrt{k\frac{1-\lambda^2}{  \D(\tilde\alpha) } }\sin\tilde\alpha \sinh\tilde\beta d\tilde\gamma \ ,
\\
& \frak{e}^3 = \sqrt{k\frac{1+\lambda }{1-\lambda} } d\alpha \ , \quad \frak{e}^4 =  \sqrt{k\frac{1-\lambda^2}{
\D(\alpha) } }\sin\alpha d\beta \ , \quad
\frak{e}^5 =  \sqrt{k\frac{1-\lambda^2}{  \D(\alpha) } }\sin\alpha \sin\beta d\gamma \ ,
\\
&
\frak{e}^{x^i} = d x^i \ ,\quad i=6 ,\dots, 9 \ ,
\end{split}
\label{djfh9}
\ea
so that the metric is then
\begin{equation}
 ds^2 = - (\frak{e}^{0})^2 +  (\frak{e}^{1})^2+ (\frak{e}^{2})^2 + (\frak{e}^{3})^2+ (\frak{e}^{4})^2+ (\frak{e}^{5})^2 + \sum_{i=6}^9 dx^i dx^i \ .
\end{equation}
Note that these simple frame fields are not the same as those defined by \ref{eq:frames1}.

Combining the $SU(2)$  result of \eqref{eq:OmegaSU2} with its $SL(2,\mathbb{R})$ counterpart,
the $\Omega$ matrix relating left and right moving frames, in this basis, has the form
\ba
&& \Omega = \frac{i}{\sqrt{\D(\a) \D (\tilde \a) } }\left(
(\lambda-1)\cos\alpha \Gamma^{345} + (\lambda + 1)  \sin\a \Gamma^3
\right)
\nonumber\\
&&\phantom{xxxxxxxxx}
\cdot\left( (\lambda-1)\cos\tilde\alpha
\Gamma^{012} + (\lambda + 1)  \sin\tilde{\a}
\Gamma^{0}  \right) \ .
\ea

Then we use our proposal for the RR fields described by eq.~\eqref{RRrule}
in this case with $\mathbb{F}_{0} = F_{3}+\star F_{3}$ and $\Phi_{0}=0$.   The new polyform
  \be
    \slashed{\mathbb{F}}   =  \mu(\lambda)  e^{-\Phi}  \slashed{\mathbb{F}}_{0} \cdot \Omega^{-1}  \ ,
 \ee
obtained in this way has components
 \begin{equation}
\begin{aligned}
&
F_1 = i \mu(\lambda) (1-\lambda^2) \left( \cos \alpha \sin\tilde\alpha \frak{e}^{0} + \cos\tilde\alpha \sin\alpha \frak{e}^3 \right)\ ,
\\
&
F_3=  i \mu(\lambda) \Big( (1- \l)^2 \cos\a\cos\tilde\alpha (\frak{e}^{012}  + \frak{e}^{345} )
-  (1+\lambda)^2\sin\a\sin\tilde\alpha (\frak{e}^{045}  +\frak{e}^{123} )   \Big)\ ,
\\
&
F_5= (1+\star) f_5\ ,
\quad f_5 = - i \mu(\lambda) \left(1-\lambda^2\right)
\Big( \sin \a \cos\tilde\a \frak{e}^{01245}
+ \cos \a \sin\tilde\a \frak{e}^{12345}\Big)\ ,
\end{aligned}
\end{equation}
with
\be
\mu(\lambda) = {4\l\ov \sqrt{k} (1-\l)^{1/2} (1+\l)^{3/2}}\ .
\ee
Then, the Bianchi identities and  equations of motion of the type-II  supergravity  are solved.   However the
fluxes are pure imaginary meaning that this should be
interpreted in the context of the type-II$^\star$ theory described in \cite{Hull:1998vg}.
This arises because the $\Omega$ matrix involves a $\Gamma^{0}$ and
therefore has similar features to performing a time-like T-duality.

In the $\lambda \rightarrow 0$ limit one immediately recovers the
geometry $AdS_3 \times S^3 \times T^4$ supported by NS flux and in
the $\lambda \rightarrow 1$ limit we find the non-Abelian T-dual of
$AdS_3 \times S^3 \times T^4$ supported by RR flux.

\section{$\lambda$-deformations for cosets }\label{sec:cosets}

In this section we will  let $\{T^A\}$ be the generators of $G$, $\{T^a\}$ be those of some subgroup $H\subset G$
and $\{T^\alpha\}$ the remaining generators for the coset $G/H$.  We will also need a second subgroup $K \subset G$ and denote its generators by $\{ T^m \}$.

The $\sigma$-model on the geometric coset $G/H$ is given by
\be\label{G/Hpcm}
S_{G/H}(\tilde g) = \frac{\kappa^2}{\pi} \int \delta_{\a \b} L_+^\a L_-^\b\ ,
\ee
where the sum is over only coset indices.  This action has a local invariance
$\tilde{g} \rightarrow \tilde{g} h$ for $h\in H$ and so depends on $\dim(G)- \dim(H)$ degrees of freedom.
We now consider the sum of this action with that of the WZW model  \eqref{eq:WZW}  and gauge a subgroup $K\subset G$ that acts as
\be
\tilde{g} \rightarrow k^{-1} \tilde{g} \ , \qq g \rightarrow k^{-1} g k \ .
\ee
We introduce a connection that transforms as
\be
A \rightarrow  k^{-1} A k - k^{-1} dk  \ ,
\ee
and repeat the analogous steps in the gauging, replacing derivatives in the PCM to covariant ones
and replacing the WZW for $G$ with a $G/K$ WZW model, giving the gauged action
\ba
&& S = S_{{\rm WZW},k}(g) +  {k\ov \pi} \int {\rm Tr}(A_-\del_+ gg^{-1} - A_+ g^{-1} \del_- g + A_- g A_+ g^{-1}- A_-A_+)
\nonumber\\
&&\phantom{xxxxxx} -  {\k^{2}\ov \pi} \int \d_{\a\b}(\tilde g^{-1}   D_+ \tilde g)^\a  (g^{-1}   D_+ \tilde g )^\b\ ,
\ea
where
\be
  D_\pm \tilde g = \del_\pm \tilde g - A_\pm \tilde g \ .
\ee
 Integrating out the $A_\pm$'s gives
\ba
&& S = S_{{\rm WZW},k}(g) -{k\ov\pi} \int [J_+ + \tilde L_+ (\l^{-1} -1) \tilde D^T]^m
\nonumber
\\
&&\phantom{xxxxxx}
[D^T-1-\tilde D(\l^{-1}-1)\tilde D^T]^{-1}_{mn} [J_- - \tilde D (\l^{-1}-1)\tilde L_-]^n\ ,
\ea
in which the $\tilde{L}$ and $\tilde D$ are the left-invariant forms
and adjoint matrix for the PCM group element $\tilde{g}$ and the contracted indices are running over the gauge group $K$.

Let us focus our attention on the case where $K = G$, i.e. we gauge the entire global $G$-symmetry.
In that case we may partially fix the  gauge by setting $\tilde{g}=\mathbb{1}$.
This will however leave a residual $H$ gauge symmetry that will be used to fix $\dim(H)$ degrees of freedom in $g$.
Then the action becomes
\be
\label{Sdefcoset}
S = S_{{\rm WZW},k}(g)  - \frac{k}{\pi} \int R_+^A  (M^{-1})_{AB} L_-^B\ ,
\ee
 where
\be
M_{AB}  = \left( \begin{array}{cc}
 (D^T - \mathbb{1})_{ab}  & (D^T)_{a\beta} \\
(D^T)_{\a b}  & ( D^T - \lambda^{-1} \mathbb{1})_{\alpha \beta }   \\
\end{array}\right) \ .
\ee
For the case where the coset is a symmetric space,\footnote{A symmetric coset  $G/H$ is one
for which the algebra $\frak{g}$ of $G$ admits a $\mathbb{Z}_{2}$ grading $\frak{g}= \frak{g}^{(0)}\oplus\frak{g}^{(1)}$
where $ \frak{g}^{(0)} = \frak{h}$ is the algebra of $H$  and with  $[  \frak{g}^{(0)} ,
\frak{g}^{(0)} ] \subset  \frak{g}^{(0)}$,  $[  \frak{g}^{(0)} ,  \frak{g}^{(1)} ] \subset  \frak{g}^{(1)}$,
$[  \frak{g}^{(1)} ,  \frak{g}^{(1)} ] \subset  \frak{g}^{(0)}$.} integrability of this theory was proved in  \cite{Hollowood:2014rla} using  the
gauged WZW-like origin of the construction of the action.
We also note that the action \eqn{Sdefcoset} with the given expression for $M_{AB}$ arises if we simply set
in \eqn{tdulalmorev2} the block diagonal part of the matrix $\l$ corresponding to the subgroup $H$ to unity.

The $\lambda \rightarrow 1$ limit, together with appropriate rescalings results in the non-Abelian T-dual
of the geometric coset as constructed in \cite{Lozano:2011kb}.
On the other hand the $\l\to 0$ limit of \eqref{Sdefcoset} gives
 \be
S= S_{{\rm WZW},k}(g) -{k\ov \pi} \int J_+^a (D^T-1)^{-1}_{ab} J_-^b  + {\cal O}(\l)\ .
\ee
which is the $\s$-model corresponding to the gauged WZW model for $G/H$ as expected. The leading correction that drives the
model away from the CFT point is proportional to
\be
\int {\rm Tr}(T^\a D^0_+ g g^{-1}) {\rm Tr}(T^\a g^{-1} D^0_- g)\ ,\qq  D_\pm^0 g = \del_\pm g -[A^0_\pm,g]\ ,
\ee
where $A^0_\pm$ are the solution for the gauge fields arising from integrating them out in the action after setting $\l=0$. Explicitly
\be
A^{0a}_+ = - i (D-\mathbb{1})^{-1}_{ab} J_+^b\ ,\qq A^{0a}_- = i (D^T-\mathbb{1})^{-1}_{ab} J_-^b\ .
\ee
This term can be written at a bilinear in the classical parafermions \cite{Bardacki:1990wj,Bardakci:1990ad} and therefore it has
a precise CFT interpretation. For the perturbation on $SU(2)/U(1)$ the above considerations were made explicit in \cite{Sfetsos:2013wia}.

One generalisation of this  construction is given by replacing the inner product $\delta_{\a \b}$
entering in \eqref{G/Hpcm} with  general metric $E_{\a \b}$, however the $G$ invariance condition
\be
f_{a\b\d} E_{\d\g} + f_{a\g\d} E_{\b\d} = 0\ ,
\ee
should still be obeyed.\footnote{When $G/H$ is symmetric  the Cartan-Killing metric restricted
to the coset is the unique $G$ invariant metric and so $E_{\a\b} \propto \d_{\a\b}$,
however for more general cosets one can find many examples where $E_{\a\b}$ is not the Cartan-Killing form,
for instance the most general $SU(3)$ invariant metric on six-dimensional coset $SU(3)/U(1)\times U(1)$ depends
on three real parameters \cite{MuellerHoissen:1987cq}.}
A second generalisation is to consider rather than a single WZW model, multiple WZW factors each at
different levels - we will give further details of this in the appendix.

To obtain the supergravity embedding we need to generalise the ansatz for RR fields described in
\eqref{RRrule} from the group to the coset.  Fortunately this has been done already in
the case of non-Abelian duality (the $\lambda\rightarrow 1$ limit of the present construction) in \cite{Lozano:2011kb}
and is easily extrapolated to the case at hand.
If we let $X^{\a}$ be $\dim(G)- \dim(H)$ local coordinates for the $\sigma$ model \eqref{Sdefcoset} then frame fields are obtained by defining
\be\label{eq:cosetframes}
\begin{aligned}
e_-^\a & = - \left(\frac{k}{2\l^2}(1- \l^2) \right)^{\frac{1}{2}} (M^{-1})^{\a B} L^B(g) \equiv {\cal N}_-^{\a\b} d X^\b \ , \\
e_+^\a &=  \left(\frac{k}{2\l^2}(1- \l^2) \right)^{\frac{1}{2}} (M^{-1})^{ B \a} R^B(g)  \equiv {\cal N}_+^{\a\b} d X^\b \ .
\end{aligned}
\ee
These are related by a Lorentz rotation $\Lambda =  {\cal N}_+{\cal N}_-^{-1}$
from which the corresponding spinor matrix $\Omega$ can be obtained and then the rule in \eqref{RRrule}
can be directly applied.  Of course, the exact form of these objects will depend on the way the residual symmetry is gauge fixed.

 \section{ $AdS_2\times S^2$ deformations}\label{sec:5}

\subsection{Deforming the $SU(2)/U(1)$ exact CFT}\label{sec:example2}

We follow this procedure and work out the action \eqn{Sdefcoset} for the case $G=SU(2)$ and $H=U(1)$.
We parametrize the group element as
 \be
g = e^{i (\phi_1 -\phi_2) \s_3/2} e^{i \om
\s_2}e^{i (\phi_1 +\phi_2) \s_3/2}\ . \ee

In both the gauged WZW and its deformation given by eq.~\eqref{Sdefcoset}, we can fix the $U(1)$ gauge redundancy by setting $\phi_2=0$ (one finds in an explicit calculation that $\phi_2$ enters the action only as a surface term).
The metric of the deformed $\s$-model \eqref{Sdefcoset} is then given by
\be
 ds^2 =k\left({1-\l\ov 1+\l} (d\om^2 +\cot^2\om d\phi^2) + {4\l\ov 1-\l^2}
(\cos\phi d\om + \sin\phi \cot\om d \phi)^2\right)\
\label{clpp22}
\ee
and zero antisymmetric tensor, where the parameter $\l$ is defined in \eqn{lee} and
have renamed $\phi_1$ by $\phi$.  The factor that will contribute to the dilaton in the supergravity embedding that will shall do is determined,
up to a constant piece, as
$ e^{-2\Phi} =\sin^2\om $. These expressions coincide with those found in \cite{Sfetsos:2013wia}.

For $\l\ll 1$, i.e. $k\ll \k^2$, the dominant term is that corresponding to the exact $SU(2)/U(1)$ coset CFT \cite{Bardacki:1990wj}.
It can moreover be shown \cite{Sfetsos:2013wia} that the extra term is a parafermion bilinear which corresponds to a relevant perturbation
since the parafermions have conformal dimension $1-1/k$. Hence these parafermions drive the $\s$-model away from the CFT point
in accordance with our general discussion above.
This perturbation has been shown to be integrable, massive and argued that in the
$k\to \infty$ limit the model flows under the renormalization group to the $O(3)$ $\s$-model \cite{Fateev:1991bv}.
This is consistent with the fact that $SU(2)/U(1)$ is a symmetric coset space.

\subsection{Deforming the $SL(2,\mathbb{R})/SO(1,1)$ and $SL(2,\mathbb{R})/U(1)$ exact CFTs}

We perform an analytic continuation in \eqn{clpp22} by sending
\be
k\to -k\ ,\qq \k\to i \k \ ,\qq \om\to -i \r\ ,
\ee
and as a result obtain a $\s$-model with metric
\be
 ds^2 =k\left({1-\l\ov 1+\l} (d\r^2 +\coth^2\r d\phi^2) + {4\l\ov 1-\l^2}
(\cos\phi d\r - \sin\phi \coth\r d \phi)^2\right)\ ,
\label{clpp2232}
\ee
and zero antisymmetric tensor and dilaton factor
$
e^{-2\Phi} = \sinh^2\r
$.
This background represents an integrable deformation of the exact $SL(2,\mathbb{R})/SO(1,1)$ coset CFT. 

Performing a further analytic continuation in \eqn{clpp2232} as
\be
\phi\to i t\ ,
\ee
we obtain a $\s$-model with metric
\be
 ds^2 =k\left({1-\l\ov 1+\l} (-\coth^2\r dt^2 + dt^2) + {4\l\ov 1-\l^2}
(\cosh t d\r + \sinh t \coth\r dt)^2\right)\ ,
\label{clpp223k}
\ee
zero antisymmetric tensor and the contribution to the dilaton factor
$
e^{-2\Phi} = \sinh^2\r
$.
This background for $\l=0$ corresponds to geometry of the exact $SL(2,\mathbb{R})/U(1))$ coset CFT.
It was globally extended and interpreted as a two-dimensional black hole in \cite{Witten:1991yr}.\footnote{In fact for $\l=0$,
the metric in \eqn{clpp223k} covers the patch of the geometry containing the black hole singularity,
i.e. region V in fig. 2 of \cite{Witten:1991yr}.
A different analytic continuation, or alternatively using vector rather than axial gauging, gives the geometry in region I of  the same figure.}
We shall return to the black hole interpretation shortly.

\subsection{Embedding to supergravity}

Consider the  ten-dimensional metric arising from combining \eqn{clpp22} and \eqn{clpp223k} with the six-dimensional flat metric on the $T^6$
\ba
&& ds^2 = k\left({1-\l\ov 1+\l} (-\coth^2\r dt^2 + d\r^2) + {4\l\ov 1-\l^2}
(\cosh t d\r  + \sinh t \coth \r d t)^2\right)
\nonumber\\
&&  \phantom{xxx} +
 k\left({1-\l\ov 1+\l} (d\om^2 +\cot^2\om d\phi^2) + {4\l\ov 1-\l^2}
(\cos\phi d\om + \sin\phi \cot\om d \phi)^2\right)
\label{met1}
\\
&&  \phantom{xxx} + \sum_{i=4}^9 dx_i^2\ .
\nonumber
\ea
In addition, by combining the corresponding dilaton factor we have for the dilaton
\be
e^{-2\Phi} = \sin^2\om \sinh^2 \r\ .
\label{fllf}
\ee
The antisymmetric tensor vanishes.
In order to satisfy the supergravity equations of motion we need to turn on flux fields. To present them
we first define the frames
\ba
\label{rfammeee}
&& e^0 = \sqrt{k{1-\l\ov 1+\l}} (\sinh t d\r + \cosh t \coth \r dt )\ ,
\nonumber\\
&& e^1 = \sqrt{k{1+\l\ov 1-\l}} (\sinh t \coth \r dt + \cosh t d \r )\ ,
\nonumber
\\
&& e^2= \sqrt{k{1-\l\ov 1+\l}} (\cos\phi \cot\om d\phi - \sin\phi d\om)\ ,
\\
&& e^3 =\sqrt{k{1+\l\ov 1-\l}} (\cos\phi d\om  + \sin\phi\cot\om d\phi)\ .
\nonumber
\ea
We will also denote by $J_2$ the Kahler form and by $J_3$ the real part of the complex differential form of type $(3,0)$ in $\mathbb{R}^6$.
In a convenient basis we have that
\be
\begin{split}
& J_2 = dx_1 \wedge dx_2 + dx_3 \wedge dx_4 + dx_5\wedge dx_6\ ,
\\
& J_3 =  dx_1 \wedge dx_3 \wedge dx_5 -  dx_1 \wedge dx_4 \wedge dx_6 -  dx_2 \wedge dx_4 \wedge dx_5 -
 dx_2 \wedge dx_3 \wedge dx_6\ .
\end{split}
\ee
The NS sector fields can be supported in a  full supergravity solution either within type-IIB or within type-IIA supergravity.

\no
Within type-IIB we have the five-form RR flux
\be
{\rm IIB}:\qq F_5 = (1+\star )f_5 \ ,\qq f_5 = {1\ov \sqrt{k}} \sqrt{4 \l\ov 1 - \l^2}
\sin\om \sinh \r\ e^0\wedge e^3\wedge J_3\ .
\ee

\no
Within type-IIA with the two- and four-form RR fluxes are
\ba
{\rm IIA}: && F_2 = {1\ov \sqrt{k}} \sqrt{4 \l\ov 1 - \l^2}
\sin\om \sinh \r\ e^0\wedge e^3\ ,
\nonumber\\
&&  F_4 = {1\ov \sqrt{k}} \sqrt{4 \l\ov 1 - \l^2}
\sin\om \sinh \r\ e^1\wedge e^2\wedge J_2\ .
\ea
These are real forms, and are solutions of type-II supergravity.
The form of the RR fields may be established using the action of  $\Omega = \Gamma^{1}\Gamma^2$ on either the
type-IIB or type-IIA embedding of the $AdS_2 \times S^2$ PCM (see \cite{Sorokin:2011rr}
for discussion of the GS action for the superstring in this background and its integrability).

\no
\subsubsection{The non-Abelian T-dual limit and the near singularity region}

Then above geometry is singular for $\r=0$ and for $\om=0$ where, for instance, the scalar curvature blows up.
We are interested in magnifying the geometry around these points. It turns out that we have to zoom in also at specific sections for the
variables $t$ and $\phi$.
Indeed after letting
\be
t={\tau \ov 2 k}\ ,\qq \r={r\ov 2k}\ ,\qq \phi={x_1\ov 2k}\ ,\qq \om= {x_2\ov 2k} \ ,\qq \l =1-{1\ov k} +\dots\ ,
\ee
we obtain that
\be
ds^2 =\ha \left( -{d\tau^2\ov r^2} + \left(dr + \tau {d\tau\ov r}\right)^2\right)+ \ha \left( {dx_1^2\ov x_2^2} + \left(dx_2
+ x_1 {dx_1\ov x_2}\right)^2\right)  + \sum_{i=4}^9 dx_i^2\
\ee
and for the dilaton
\be
e^{-2\Phi} = x_2^2 r^2\ .
\ee
Compared with \eqn{fllf} we have shifted the dilaton so that $e^\Phi$ gets multiplied by $4k^2$ and $\Phi$ remains finite in the
above limit.
That implies that the fluxes are also multiplied by $4 k^2$ so that the Einstein equation of motion is satisfied.
The result for this limiting procedure for the fluxes of type-II supergravity gives
\be
{\rm IIB}:\qq  F_5 = (1+\star )f_5 \ ,\qq f_5 = {1\ov \sqrt{2}}  \ d\tau \wedge (x_2 dx_2 + x_1 dx_1)\wedge J_3\
\ee
 and
\ba
{\rm IIA}: && F_2 = {1\ov \sqrt{2}}  \ d\tau \wedge (x_2 dx_2 + x_1 dx_1)\ ,
\nonumber\\
&& F_4 =  {1\ov \sqrt{2}} (r dr + \tau d\tau )\wedge dx_1\wedge J_2\ .
\ea
In conclusion the non-Abelian T-dual in this case provides the geometry near the singularities
when the parameter $\l$ tends to unity.

\subsection{Global structure}

In this section we will study the geometry presented here in the context of the two-dimensional black hole
solution of \cite{Witten:1991yr}.
First let us just consider the two-dimensional metric \eqref{clpp223k} and obtain its conformally
flat form by making a coordinate transformation
\be
u = \cosh \r (e^{-t} + \l  e^{t})\ ,\qq v = \cosh \r (e^{t} + \l  e^{-t})\ .
\label{uvhf}
\ee
Then \eqref{clpp223k} becomes
\be
ds^2 = k (1-\l^2) {du dv\ov f(u, v)}\ , \quad f(u,v)= (u-\l v)(v-\l u) - (1-\l^2)^2 \ .
\label{clppuv}
\ee
For $\lambda =0 $ this coincides with the global metric found in eq. (28) of  \cite{Witten:1991yr}.
At face value the effect of the $\l$-deformation is to modify the location of the singularity defined by $f(u,v)= 0$.
This is illustrated in the Penrose diagram  of fig. 1.
A peculiar feature is that the deformed black-hole singularity curve is no longer a horizontal line extending to null
infinities ${\cal I}^\pm$ but instead ``bends'' back on itself in the Penrose diagram to close off into a tear drop shaped
ending at future time-like infinity.  As a consequence it appears that   some portion of the singularity is not protected by a horizon.
This is clearly a puzzling feature and warrants a further study. One likely resolution is that this portion of the space time should be excised.
The study of this metric is made somewhat more difficult since for $\l \neq 0,1,$ it does not admit
any isometries as can be seen by direct inspection of the Killing equations.

\vskip 0 cm
\begin{figure}[!h]
\label{fig:penrose}
\begin{center}
 \includegraphics[height=6cm]{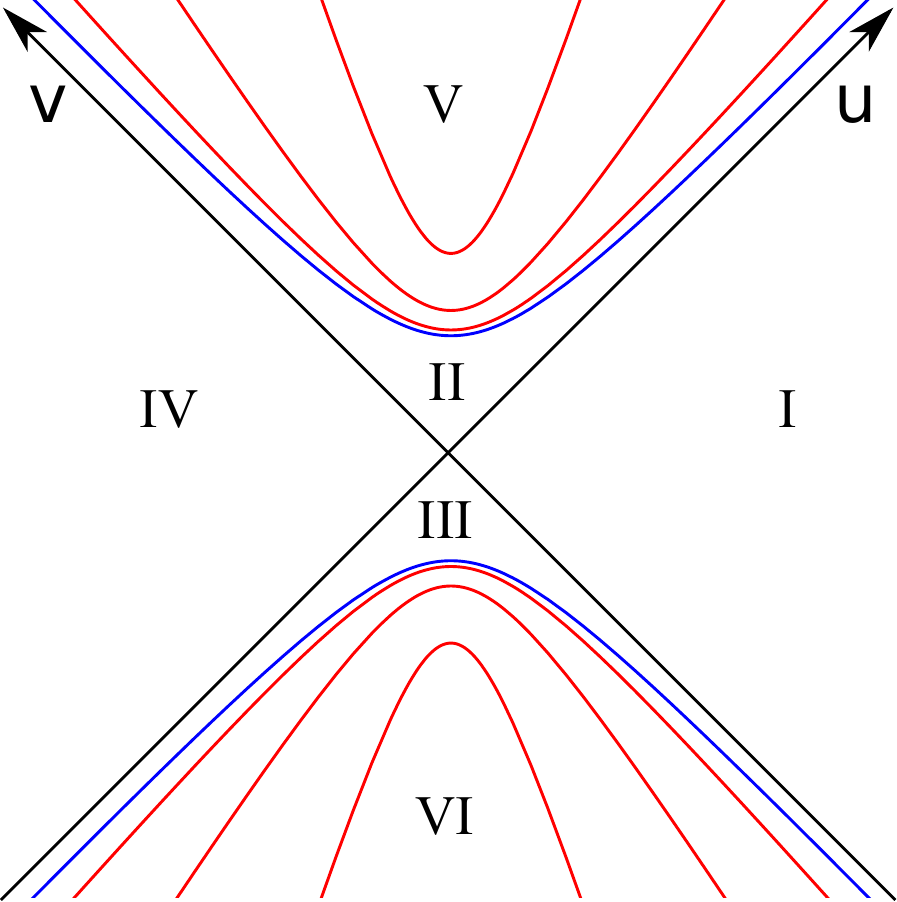}
\hskip 2 cm
\includegraphics[height=6cm]{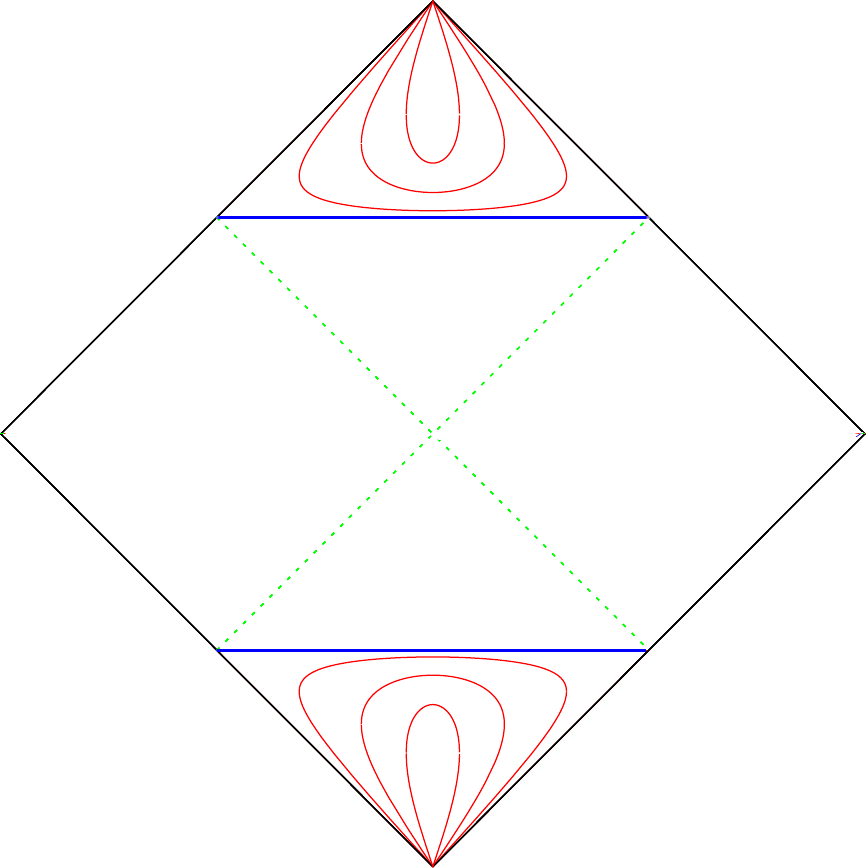}
 \end{center}
\caption{\footnotesize{ Kruskal (left) and Penrose (right) diagram of deformed space time showing
how the location of singularity (red lines) migrates.  Shown in blue is the undeformed $\l = 0$
singularity at $uv =1 $ and its corresponding horizon is displayed on the Penrose diagram in green.  In red are the singular curves corresponding to $f(u,v)=0$ for $\l=(0.05,0.2,0.5)$ which forming increasingly sharper tear drop regions in the Penrose diagram  as  $\l \to 1$.   }}
\end{figure}
It is interesting to consider the limit as $\l \to 1$ where in the Penrose diagram the singularities curves degenerate into vertical lines.
As stated above, if the coordinates are also scaled in this limit one obtains a non-Abelian T-dual geometry that probes the singularity.
Instead let us take a different limit under which the coordinates are not scaled. We simply let $\l\to 1$ by sending $ k\to \infty$ with $\kappa^2$ fixed.   Then the metric \eqref{clppuv} becomes
\be
ds^2 = -k (1-\l^2) \frac{du dv}{(u-v)^2} + \cdots  = \k^2 {dz^2 - dt^2 \ov z^2}+ \cdots \ ,
\ee
so that an $AdS_2$ geometry of size $\k^2$ emerges. Hence,
taking this limit in this way one (necessarily) shows that the singularity is removed.

We now turn to a second question: can this full globally extended geometry be supported in type II supergravity?
In fact, the answer is no and the reason can be seen even in the context of the undeformed black-hole for $\l = 0$.
In this   case the singularity lies on $uv=1$ and in crossing the singularity (going from region V to  II
in the terminology of \cite{Witten:1991yr}) the dilaton $\Phi =-\ha \ln (1- uv)$  picks up a shift of $i \pi$.  Of course, when $\l=0$ the geometry
does not requires any RR flux and so the additive shift to the dilaton can be, to an extent, ignored at the level of solving the supergravity equations.
However when $\l\neq 0$, if one wishes to keep the dilaton real,  this will necessitate allowing the RR fluxes to become imaginary.
Thus there is no global extension that can cover  both asymptotic regions whilst keep the solution in type II.

\section{An $AdS_3 \times S^3$ coset deformation}
\label{sec:example3}

\subsection{ Deforming the $SO(4)/SO(3)$ coset}
 We parametrize the group $SO(4)$ element $g$ in the  $SU(2)\times SU(2)$ decomposition with
 \be
 g_1 = \left( \begin{array}{cc}
  \alpha_0 + i \alpha_3 & \a_2 + i \a_1   \\
 -\a_2 + i \a_1  &   \a_0 - i \a_3   \\
\end{array}\right)  \ , \quad
   g_2 = \left( \begin{array}{cc}
  \beta_0 + i \b_3 & \b_2 + i \b_1   \\
 -\b_2 + i \b_1  &   \b_0 - i \b_3   \\
\end{array}\right) \ ,
 \ee
 with the usual determinant constraints.
 A nice gauge   choice that completely fixes the residual $H= SU(2)$ symmetry is
 \be
 \a_2 = \a_3 = \b_3 = 0
 \ee
and $H$ invariant combinations  of the remaining coordinates are given by
\be
\alpha = \alpha_1 = (1- \alpha_0^2)^\frac{1}{2}  \ , \quad  \gamma = \beta_1 \alpha_1 \ ,  \quad   \beta = \left(\beta_1^2 + \beta_2^2 \right)^\frac{1}{2} \ ,
\ee
which are simply $|\vec{\a}|, |\vec{\b}|$ and $\vec{\a}\cdot\vec{\b}$ after gauge fixing (in what follows we prefer
to use the invariants $\a_0$ and $\b_0$ rather than $\a$ and $\b$ since they lead to marginally more concise expressions).

Following the procedure described in section \ref{sec:cosets} results in an effective target space  geometry
\ba
\label{dsdef}
&& ds^2  = \frac{k}{ 2(1-\l^2)\L} \Big( \Delta_{\a\a} d\a_0^2 + \Delta_{\b\b} d\b_0^2 + \Delta_{\g \g} d\g^2
\nonumber\\
&& \phantom{xxxxx}
+ 2 \D_{\a \b}d\a_0 d\b_0 + 2\Delta_{\a\gamma} d\a_0 d\g    + 2\Delta_{\b\gamma} d\b_0 d\g \Big)\ ,
\ea
with
\be
 \Lambda =(1-\a_0^2)(1-\b_0^2) - \g^2
\label{lambbb}
\ee
and
\ba
&&
\Delta_{\a\a} =  4 (1+\l)^2 - \b_0^2 (3+\l)(1+3\l) \ ,
\nonumber\\
&&
\Delta_{\b\b}  =  4 (1+\l)^2 - \a_0^2 (3+\l)(1+3\l)\ ,
\nonumber
\\
&& \Delta_{\g\g} =    (1-\l)^2\ ,
\\
&& \Delta_{\a \b} =  \a_0 \b_0 (1-\l)^2 + 4 \g (1+\l)^2
\nonumber\\
&& \Delta_{\a\g}  = -\b_0(1-\l)^2\ , \qq \Delta_{\b\g}  = -\a_0(1-\l)^2 \ .
\nonumber
\ea
The NS two-form potential can be chosen to be zero and the contribution to the dilaton factor is given (up to a constant shift) by
$
e^{-2\Phi} = \L
$.
The contribution to the dilaton beta function equation turns out to be $\displaystyle \frac{6}{k} \frac{1+ \l^2} {1- \l^2}$ which is
just a constant. This is an important consistency check that there is a possibility to embed this into a full supergravity solution by combining
with another model that may contribute exactly the opposite.

\no
Note the ranges of the coordinates
\be
\label{eq:domains}
0< \a_0^2 <1 \ , \quad 0< \beta_0^2 <1 \ , \quad \g^2 < (1- \a_0^2)(1-\b_0^2)  \ ,
\ee
ensure that  on this domain the function $\Lambda \geqslant 0$.
These are important to keep in mind since they ensure that the metric has positive signature (that is to say if one uses the metric
\eqref{dsdef} blindly and goes beyond the range  of  these coordinates, it  does not remain positive).
 There is a manifest $\mathbb{Z}_2$ symmetry  swapping $\alpha_0$ and $\beta_0$
just corresponds to a switching of the two $SU(2)$ factors of the $SO(4)$ decomposition.

In the limit $\lambda \rightarrow 0$ the geometry is precisely that corresponding to the $G/H$ gauged WZW model
written explicitly in this parametrisation in (4.12) (for $r=1$) of \cite{Polychronakos:2010fg}.

The limit $\lambda \rightarrow 1$ needs to be taken with some care. Defining
\be
\lambda = \frac{k}{k + \kappa^2} \ , \qq \a^2= -\frac{ t_1}{k^2} \ ,  \qq  \b^2 = -\frac{t_3}{k^2} \ , \qq  \g = \frac{t_2}{k^2} \
\ee
and taking the limit $k\rightarrow \infty$ one recovers (setting $\kappa = \sqrt{2}$)
the metric of the non-Abelian T-dual of $G/H=S^3$ with respect to $G=SO(4)$ obtained in (4.20) of \cite{Lozano:2011kb}.

The frame fields for the geometry are defined by eq.~\eqref{eq:cosetframes} and
are given\footnote{To present these frames in a way that is deomocratic between $\a_0$ and $\b_0$ we in fact
perform a supplementary rotation in the $4-5$ plane on those frames given by a direct application of \eqref{eq:cosetframes}.}
by (we start the frame numbering at $3$ for reasons that become obvious momentarily):
  \def\C{{\cal C}}
 \def\S{{\cal S}}
 \be
\begin{aligned}\label{eq:frameS3coset}
e_\pm^{3} &=   \sqrt{{k\ov 2\L} {1-\l\ov 1+\l}}
\left(\b_0 d\a_0 + \a_0 d\b_0 - d\g  \right) \ ,
 \\
e_\pm^{4} &=  \pm  \sqrt{{2 k\ov \L} {1+\l\ov 1-\l}} \sin{\psi\ov 2}
\left( \sqrt{1- \b_0^2} \ d\a_0 - \sqrt{1- \a_0^2}\  d\b_0 \right) \ ,    \\
e_\pm^{5} &=    \mp   \sqrt{{2 k\ov \L} {1+\l\ov 1-\l}} \cos{\psi\ov 2}
 \left(\sqrt{1- \b_0^2} \ d\a_0 + \sqrt{1- \a_0^2}\ d\b_0 \right)  \ ,
\end{aligned}
\ee
where the angle $\psi$ is defined through
\be
\label{eq:defpsi}
\g  = \sqrt{(1- \a_0^2)(1- \b_0^2)}  \cos \psi   \ .
\ee
Because the plus and minus frames differ only by reflection in the $e^{4}$ and $e^{5}$
directions it is evident that the corresponding Lorentz rotation in the spinor representation will have the form $\Omega \sim \Gamma^{4}\Gamma^{5} $.

\subsubsection{Analytic continuation}
\label{sec:SO4analyticcontinuation}

Results for $(SL(2, \mathbb{R} )\times SL(2,\mathbb{R} ) ) / SL(2,\mathbb{R} )$ can be obtained by analytic continuation.
There may be several different ways to perform an analytic continuation and these are given essentially
by changing the domains of $\a_{0}, \b_{0}$ and $\g$ in eq.~\eqref{eq:domains}.  The continuation we seek should be such that the frame $e^{3}$ defined
in \eqref{eq:frameS3coset} becomes time-like and $e^{4}$ and $e^{5}$ remain space-like.
 With such a choice, the $\Omega$ matrix  will contain reflections only in space-like directions
and the resulting RR fields will remain real (other choices could lead to solutions of type II$^{\star}$).

Let $\tilde\a_{0}, \tilde{\b}_{0}$ and $\tilde \g$ be the coordinates for the analytically continued geometry
which is given by the same metric as \eqref{dsdef} but with the replacement of $k \rightarrow - k$ and domains
 \be
\label{eq:domains}
1< \tilde\a_0^2  \ , \qq 1<  \tilde\beta_0^2   \ , \qq \tilde\g^2 < (1- \tilde\a_0^2)(1-\tilde\b_0^2)  \ .
\ee
We define the function
\be
\label{eq:tildeLambda}
\tilde{\L} = (\tilde \a_0^2-1)(\tilde \b_0^2-1) - \tilde \g^2 \ ,
\ee
which is positive over its domain and the angle $\tilde\psi$ through
\be
\tilde\g  = \sqrt{(\tilde \a_0^2-1) (\tilde \b_0^2-1)}  \cos \tilde \psi \ .
\ee
The analytically continued versions of  the frame fields in \eqref{eq:frameS3coset} are given by
 \def\Ch{{\rm Ch}}
 \def\Sh{{\rm Sh}}
 \be
\begin{aligned}\label{eq:frameAdS3coset}
e_\pm^{0} &=   \sqrt{{k\ov 2\L} {1-\l\ov 1+\l}}
\left(\tilde\b_0 d\tilde\a_0 + \tilde\a_0 d\tilde\b_0 - d\tilde\g  \right) \ ,
 \\
e_\pm^{1} &=  \pm \sqrt{{2 k\ov \L} {1+\l\ov 1-\l}} \cos{\tilde \psi\ov 2}
\left( \sqrt{\tilde\b_0^2-1} \ d\tilde\a_0 - \sqrt{\tilde\a_0^2-1}\  d\tilde\b_0 \right) \ ,    \\
e_\pm^{2} &=    \mp   \sqrt{{2 k\ov \L} {1+\l\ov 1-\l}} \sin{\tilde \psi\ov 2}
 \left(\sqrt{ \tilde\b_0^2-1} \ d\tilde\a_0 + \sqrt{\tilde\a_0^2-1}\ d\tilde\b_0 \right)  \ ,
\end{aligned}
\ee
where the flat metric has signature $(-++)$. The Lorentz rotation spinor matrix is this space is given by $\tilde\Omega \sim \G^{1}\G^{2}$.

\subsection{Supergravity embedding}
\label{sec:SO4sugra}
We introduce frame fields in the $T^4$ directions $e^i = dx^i$, $i=6\dots 9$ such that the full ten-dimensional metric is
$
ds^2 = \eta_{ij} e^i e^j
$
with $e^{0}, e^1,e^{2}$ given by \eqref{eq:frameAdS3coset} and $e^{3},e^4 e^{5}$ by \eqref{eq:frameS3coset}. The dilaton is given by
$
e^{-2\Phi} = \L \tilde{\L}
$
and the NS two-form potential can be taken to be  zero.

It is evident that because we have sent $k\to - k$ in the analytic continuation,
the contributions to the dilaton equation from the $3,4,5$ directions will cancel exactly with those in the $0,1,2$ directions.
To satisfy the Einstein's equations we need simply an RR three form
\be
F_3 = 2  \sqrt{\frac{2\l\L\tilde \L }{k(1-\l^2)}}  \left(e^{045} + e^{123}  \right)\ ,
\ee
which is in keeping with the ansatz for RR fluxes given by \eqref{RRrule}
when the spinor rotation matrix is the combination of that in the $3,4,5$ and $0,1,2$ directions i.e. $\Omega \sim \G^{12} \G^{45}$.
This flux also solves its Bianchi identity and equation of motion.
One can see that all components of $F_3$ are extended in either or both the non-compact directions $\tilde\a_0$ and $\tilde\b_0$,
for this reason there seems to be no well defined (i.e. finite) charge associated to this flux.   A final comment is that since this background only has an $F_3$ active, the S-dual will have purely NS flux which may be useful for further investigation.

 \section{Conclusions and discussion}

 In this work we have demonstrated very explicitly that the  NS of backgrounds corresponding to integrable $\lambda$-deformations
can be upgraded to full solutions of supergravity supported by appropriate Ramond fluxes.  This gives very convincing support that
the $\lambda$-deformation of a supercoset $\s$-model (such as that for the $AdS_5 \times S^5$ string) will be an exactly marginal
deformation and correspond to the $q$ root of unity quantum deformations as postulated in \cite{Hollowood:2014qma}.
Extracting the supergravity background in the specific case of $AdS_5\times S^5$ will be the subject of future work.

The examples presented within preserve no isometries and are thus very unlikely to be  supersymmetric.
Also the defomations act equivalently in the $AdS$ and sphere parts of spacetime.  It would be extremely interesting if one finds
a way to avoid either of these features; preserving even ${\cal N}=1$ supersymmetry would be desirable and  acting just in
the sphere directions would result in a deformation to the geometry for which the dual field theory would remain conformal.
If this is the case one might be able to understand more clearly the consequences of these $\l$-deformations for holography.

In this work we have been implicitly always thinking of the $\l$-deformation as being a deformation away from the CFT defined by a (gauged)-WZW model.
This is reflected in the fact that our gauge fixing choice   always  involved fixing the group element defining the PCM $\tilde{g}=\mathbb{1}$.
One consequence of this is that geometrically it is hard in general to see the PCM geometry emerging (it is the non-Abelian T-dual of the
PCM that is recovered in a limit in which coordinates are also scaled as $\l \to 1$). 
 Prior to gauge fixing the PCM and the WZW model are treated on equal footings.  One may thus like to reconsider the system instead
as deformation away from the PCM point.  One might expect that this can be done by adopting a different gauge fixing choice in which the group element
defining the PCM is left untouched.  In support of this, we saw for the case of $Sl(2)/U(1)$ that by making an appropriate coordinate transformation an $AdS_{2}$ space did indeed emerge in the $\l \to 1$ limit.   There are however two difficulties with this, firstly it is not possible to completely fix
the gauge symmetry in this way and secondly that experience dictates that different
gauge fixing choices do no more than generating diffeomorphisms of the target space.      This issue warrants further consideration.

 \section*{Acknowledgements}
We thank B. Craps, S. Cremonesi, J. Russo, A. Sevrin and J. Vanhoof for helpful discussions.
The research of K. Sfetsos is implemented
under the \textsl{ARISTEIA} action (D.654 of GGET) of the \textsl{operational
programme education and lifelong learning} and is co-funded by the
European Social Fund (ESF) and National Resources (2007-2013).
 The work of DCT was supported in part by FWO-Vlaanderen through project G020714N
and postdoctoral mandate 12D1215N, by the Belgian Federal Science Policy Office through the Interuniversity Attraction Pole P7/37,
and by the Vrije Universiteit Brussel through the Strategic Research Program ``High-Energy Physics''.

\newpage

\appendix

\section{$\prod_{i=1}^N G_{k_i}/G_{\sum_{i=1}^N k_i}$}

We may easily generalize the construction for the case of direct groups.   Consider a WZW model on  $G_{k_1}\times G_{k_2}\times \cdots \times G_{k_N}$ and we gauge the $G$ action of each of these factors to give a  copies of a $G/G$ WZW at all the levels.  In addition we consider a PCM on a coset $\prod_{i=1}^N G_i/ H $ where $H$ is a diagonal subgroup of $G$.   We find it convenient to work with a block diagonal realization of this set up
 \be
T^{\bar{A}}= T^A_{i}= {\rm diag}(0,0,\dots ,\underbrace{t^A}_{i{\rm th}} ,0,\cdots , 0)\ , \qq i =1,2,\dots , N\ .
 \ee
in which in introduce the composite index $\bar{A} = A i$ . The subgroup $H$  is  generated by $t^a = (t^a,t^a,\dots , t^a)$ and then the coset is comprised by all generators $T^{\bar{A}}$, except these. Then the group element is $g=(g_1,g_2,\dots , g_N)$, where $g_i \in G_i$ and the gauged WZW is given by the usual formulas but with the modification that the inner product is normalised such that $k_i$ appears for each block.  As before we consider the case in which group element of the PCM on the coset is fixed to $\tilde g = \mathbb{1}$.    Then all the formulae of the previous section can be applied directly.

\subsection*{ Example: $SU(2)_{k_1} \times SU(2)_{k_2}$}

Consider the simplest case of an $SU(2)_{k_1} \times SU(2)_{k_2}$ WZW model.  Let  $ \sigma^A$ be the generators of each $SU(2)$ block such that
\be
T^{\bar A}  = \left\{ \begin{array}{c}  {\rm diag}( \s^A, 0)  \ , \quad \bar{A}= 1\dots 3 \\  {\rm diag}(0, \s^A )  \ , \quad \bar{A}= 4\dots 6 \end{array} \right. \ ,
\ee
  with the subgroup $H$ being generated by ${\rm diag} (\s^A, \s^A)$.

Then one finds an action after gauge fixing the PCM model given by
\be
S_{tot} = k_1 S_{WZW}[g_1] +  k_2 S_{WZW}[g_2]  + \frac{k}{\pi} \int i A_-^{\bar A} J_+^{\bar A} - i A_+^{\bar A} J_-^{\bar A} +  A_+^{\bar A} M_{\bar{A}\bar{B} }A_-^{\bar A} \ ,
\ee
where $k= k_1+ k_2$,  $s_i = k_i/k$ and
\be
J_\pm^{\bar{A} } = \left\{ \begin{array}{c}  s_1 J_\pm^A[g_1]    \\  s_2 J_\pm^A[g_2]     \end{array} \right. \ , \quad M_{\bar{A}\bar{B} } = \left\{ \begin{array}{c|c }  s_1 D_{BA}(g_1) + \left(  \frac{\k^2}{ k}- s_1\right)
\delta_{AB}  & -\frac{\k^2}{ k}
\delta_{AB} \\ \hline - \frac{\k^2}{ k}
\delta_{AB}   & s_2 D_{BA}(g_2) + \left(  \frac{\k^2}{ k}- s_2\right) \end{array} \right\}   \ .
\ee

We make the gauge fixing exactly as in section \ref{sec:example3} and find after integrating out the gauge fields a $\s$-model on a target space
\ba
\label{dsdefk1k2}
&& ds^2  = \frac{k_1+k_2}{(1-\l)\Lambda} \Big( \Omega_{\a\a} d\a_0^2  +  \Omega_{\b\b} d\b_0^2+   \Omega_{\g \g} d\g^2
\nonumber\\
&&\phantom{xxxxx}
 +2 \Omega_{\a \b}d\a_0 d\b_0 + 2 \Omega_{\a\gamma} d\a_0 d\g    + 2 \Omega_{\b\gamma} d\b_0 d\g  \Big)\ ,
\ea
with
\ba
 && \Omega_{\a\a} =   (1+ r)^{-2} Z^{-1} \left[ Z^2 - \left(Z^2-  (1-\l)^2(1+r^{-1})^2 \right)\b_0^2 \right]\ ,
 \nonumber
 \\ && \Omega_{\b\b}  =  (1+ r^{-1})^{-2} Z^{-1} \left[ Z^2 - \left(Z^2-  (1-\l)^2(1+r)^2\right)\a_0^2 \right]\ ,
 \nonumber
 \\
&&  \Omega_{\g\g} =    (1-\l)^{2} Z^{-1}\ ,
\\
&&
\Omega_{\a \b} =  (1-\l)^2 Z^{-1} \a_0 \b_0 + r (1+r)^{-2} Z \g
\nonumber\\
&&
 \Omega_{\a\g}   = -r^{-1} (1-\l)^2 Z^{-1} \b_0\ , \quad
 \Omega_{\b\g}   = -r (1-\l)^2 Z^{-1} \a_0 \ ,
\nonumber
\ea
where
\be
r = \frac{k_2}{k_1} \ ,\qq Z =  8 \l + (1-\l) r^{-1}(1+r)^2 \ .
\ee
For the case of $r=1$ i.e. equal levels, this metric reduces to that of Section \ref{sec:example3}
and for unequal levels but with the deformation turned off (i.e. $\l=0$) they give the geometry \cite{Polychronakos:2010fg},
i.e. eq. (4.12) (see also \cite{Crescimanno:1991bz}) .  There is a also a manifest $\mathbb{Z}_2$
invariance under exchange of $\a \leftrightarrow  \b$ and $k_1\leftrightarrow  k_2 $.

Frame fields for this geometry are found using the general formula \eqref{eq:cosetframes} and are given by
  \def\C{{\cal C}}
 \def\S{{\cal S}}
 \be
\begin{aligned}
\label{eq:frameS3cosetK1K2}
e_\pm^{3} &=  \sqrt{(k_1+k_2) (1-\l)\ov \L Z}
\left(r^{-1}\b_0 d\a_0 + r \a_0 d\b_0 - d\g  \right) \ ,  \\
e_\pm^{4} &=  \pm  \sqrt{{k_1k_2\ov k_1+k_2} {Z\ov (1-\l) \L}}
\sin \frac{\psi}{2} \left(r^{-1/2} (1- \b_0^2)^\frac{1}{2}   d\a_0 - r^{1/2}  (1- \a_0^2)^\frac{1}{2}  d\b_0 \right) \ ,    \\
e_\pm^{5} &=    \mp  \sqrt{{k_1k_2\ov k_1+k_2} {Z\ov (1-\l) \L}}
\cos \frac{\psi}{2} \left(r^{-1/2} (1- \b_0^2)^\frac{1}{2}  d\a_0 + r^{1/2}  (1- \a_0^2)^\frac{1}{2} d\b_0 \right)  \ ,
\end{aligned}
\ee
where  the angle $\psi$ is given by \eqref{eq:defpsi}.
The NS two form potential can be chosen to be zero and the dilaton is given (up to a constant shift) by  $e^{-2\Phi} = \L $.
The dilaton beta function equation gives a constant as is required for this to be embedded into a full supergravity solution.

 \subsection*{IIB embedding}
 The $SL(2)_{k_1} \times SL(2)_{k_2}$ result can be obtained by analytic continuation exactly as described
in section \ref{sec:SO4analyticcontinuation}; we change the domain of $\a_0$ and $\b_0$ and simultaneously flip the sign of the levels as
$k_i \rightarrow - k_i$.  Let us denote the corresponding frame fields obtained in this way as $e^0, e^1$ and $e^3$.
These will be given by the expressions in \eqn{eq:frameS3cosetK1K2} with all quantities replaced by the tilded counterparts and
with the arguments in the square roots also flipping signs, i.e. $\sqrt{\tilde \a^2-1}$.
A ten-dimensional metric is completely by appending a $T^4$ to the six-dimensional space  obtained from the $SL(2,\IR)$ and $SU(2)$ constructions.
The dilaton is then given by
 \be
e^{-2\Phi} = \L \tilde{\L}  \ ,
\ee
where $\tilde{\L}$ is as defined in \eqref{eq:tildeLambda}.  The dilaton supergravity equation
is solved by construction due to the cancelation between $SL(2)$ and $SU(2)$ factors.
The Einstein equation is solved when the geometry is supported by three-form
\be
F_3 = \mu \sqrt{\L\tilde \L} \left( e^{045} + e^{123}  \right) \ ,
\ee
where
\be
\mu^2 = \frac{1}{ k_1 r^4 Z^3N^3} \left(  1 + r^4 (1 - Z^2 N^2)^2 -2 r^2 (1+ Z^2 N^2) \right) \ ,\quad N^{-1}=(1+r)(1-\l)\ .
\ee
One may check that the constant $\m$ is indeed invariant under the above $\mathbb{Z}_2$ symmetry, albeit not manifestly.
This flux solves its Bianchi identity and equation of motion.

\end{document}